\documentclass[onecolumn]{revtex4}

\usepackage{graphicx}
\usepackage{dcolumn}
\usepackage{bm}
\usepackage{color}
\usepackage{amssymb,amsmath}

\input epsf

\begin{document}

\title{\Large{Entropy-Corrected New Agegraphic Dark Energy Model in Ho$\check{\text r}$ava-Lifshitz
Gravity}}

\author{\bf Piyali Bagchi
Khatua$^1$\footnote{piyali.bagchi@yahoo.co.in}, Shuvendu
Chakraborty$^2$\footnote{shuvendu.chakraborty@gmail.com} and Ujjal
Debnath$^3$\footnote{ujjaldebnath@yahoo.com}}
\affiliation{$^1$Department of CSE, Netaji Subhas Engineering
College, Garia, Kolkata-700 152, India.\\
$^2$Department of Mathematics,Seacom Engineering College, Howrah,
711 302, India.\\
$^3${Department of Mathematics, Bengal Engineering and Science
University, Shibpur, Howrah-711 103, India.} }

\date{\today}

\begin{abstract}
In this work, we have considered the entropy-corrected new
agegraphic dark energy (ECNADE) model in Ho$\check{\text
r}$ava-Lifshitz gravity in FRW universe. We have discussed the
correspondence between ECNADE and other dark energy models such as
DBI-essence,Yang-Mills dark energy, Chameleon field, Non-linear
electrodynamics field and hessence dark energy in the context of
Ho$\check{\text r}$ava-Lifshitz gravity and reconstructed the
potentials and the dynamics of the scalar field theory which
describe the ECNADE.
\end{abstract}

\maketitle

\section{\normalsize\bf{Introduction}}

Recent observational data indicates that our universe is currently
under accelerating phase [1 - 3]. On the basis of this observation
one can state that if Einstein's theory of gravity is acceptable
on cosmological scales, then our universe must dominated by a
mysterious form of energy called dark energy. Scalar field models
arise in string theory and are studied as promising choices for
dark energy. A lot of works dealing with scalar field dark energy
models are available which influence in the acceleration of the
universe [4-28]. Recently, the cosmic acceleration is understood
by imposing a concept of modification of gravity for an
alternative candidate of dark energy [29, 30]. This concept
provides very natural gravitational alternative for exotic matter.
This type of gravity is predicted by string/M-theory. The
explanation of the phantom or non-phantom or quintom phase of the
universe can be described by this gravity without introducing
negative kinetic term of dark energies. Recently Ho$\check{\text
r}$ava [31 - 34] proposed a new theory of gravity. It is
renormalizable with higher spatial derivatives in four dimensions
which reduces to Einstein's gravity with non-vanishing
cosmological constant in IR but with improved UV behaviours. It is
similar to a scalar field theory of Lifshitz [35] in which the
time dimension has weight 3 if a space dimension has weight 1 and
this theory is called Ho$\check{\text r}$ava-Lifshitz gravity.
Ho$\check{\text r}$ava-Lifshitz gravity has been studied and
extended in detail and applied as a
cosmological framework of the universe [36 - 38].\\

To understand the recent cosmic acceleration many theoretical
attempts have been taken in the framework of fundamental theory
such as string theory or quantum gravity. After several attempts
on some basic quantum gravitational principle scientists found two
important models named holographic dark energy model (HDE) and the
agegraphic dark energy model (ADE). HDE and ADE are originated or
motivated by holographic hypothesis [39] and uncertainty relation
of quantum mechanics. From quantum field theory, it can be
conclude that the ultraviolet cut-off $\Lambda$ could be related
to the infrared cut-off $L$ due to the limit set by forming a
black hole, obeying the fact that the quantum zero-point energy of
a system with size $L$ should not exceed the mass of a black hole,
i.e. $L^{3}\Lambda^{3}\leqslant(M_{p}L)^{3/2}$ [40]. Considering
the energy density corresponding to the zero point energy and
cut-off $\Lambda$ as $\rho_{D}\varpropto \Lambda^{4}$ the previous
expression becomes $L^{3}\rho_{D}\leqslant L M_{p}^{2}$. So the
energy density of HDE becomes
$\left(\rho_{D}=3n^{2}M_{p}^{2}/L^{2}\right)$, where $3n^{2}$ is a
constant and attached for convenience. In HDE model the black hole
entropy $S$ plays a crucial role as the HDE density depends on the
entropy area relation $S \varpropto A \varpropto L^{2}$ of black
holes in Einstein's gravity, where $A$ is the area of the horizon
[41]. If $L$ is taken as the Hubble horizon $H^{-1}$, the
observational density of the dark energy is very much comparable
to $\rho_{D}$ [42]. But there is a theoretical problem
 regarding equation of state to accelerate the universe [43].\\

Motivated from the thermal equilibrium fluctuations and quantum
fluctuations of loop quantum gravity (LQG) one can modified this
after including the logarithmic correction as  [44]

 \begin{equation}
 S=\frac{A}{4G}+\gamma \ln\frac{A}{4G}+\beta
 \end{equation}
 where $\gamma$ and $\beta$ are two dimensionless constants of
 order unity. Using this corrected entropy area relation, the
 energy density of the entropy corrected HDE can be written as [45]

 \begin{equation}
 \rho_{D}=3n^2 m_{p}^2 L^{-2}+\gamma \ln(m_{p}^2
 L^{2})L^{-4}+\beta L^{-4}
 \end{equation}
 For large $L$ equation (2) reduces to ordinary HDE energy density
 as the last two terms become negligible. So the correction make
 sense only at the early stage of the universe and when universe
 become large entropy corrected HDE reduces to ordinary HDE.\\

Since the origin of HDE and ADE are same, we would like to
 consider here the so-called Entropy Corrected New Agegraphic DE
 (ECNADE) model. In this model it is assumed that the
observed dark energy comes from the space-time and matter field
fluctuations in the universe [46, 47]. In agegraphic model, the
age of the universe is chosen as the length measure, instead of
horizon distance to avoid the casualty problem in holographic dark
energy. Cai  proposed the original ADE model to explain the
accelerated expansion of the universe on the basis of two
arguments- Karolyhazy relation [48] and Maziashvili arguments
[49]. But the original ADE model has some difficulties to explain
matter dominated epoch as the time scale is chosen as the age of
the universe. Then Wei and Cai [47] consider the time scale as
conformal time and propose the energy density of the NADE as
$\left(\rho_{x}=3n^{2}M_{p}^{2}/T^{2}\right)$ where $T$ is the
conformal time of FRW model can be written as,
\begin{equation}
T=\int\frac{dt}{a}=\int_{0}^{a}\frac{da}{Ha^2}
\end{equation}

where $a$ is the scale factor and $H\equiv\frac{\dot{a}}{a}$ is
the Hubble parameter.\\

So now one can consider just replacing $L$ of the equation (2) by
the conformal time $T$, the energy density of ECNADE as [50]

\begin{equation}
\rho_{x}=\frac{3n^2 m_{p}^2}{T^{2}}+\frac{\gamma \ln(m_{p}^2
T^{2})}{T^{4}}+\frac{\beta}{T^{4}}
\end{equation}

Similar as ECHDE we see that the last two terms of the equation
(4) can be comparable to the first term at the early epoch of the
universe and negligible when the conformal time scale $T$ becomes
large. Therefore, the entropy correction can be important at the
early time and the ECNADE model as ECHDE.\\

In this work we discuss the correspondence between ECNADE and
other dark energy models such as DBI-essence, Yang-Mills dark
energy, Chameleon field, Non-linear electrodynamics field and
hessence dark energy in the context of Ho$\check{\text
r}$ava-Lifshitz gravity and reconstructed the potentials and the
dynamics of the scalar field theory which describe the ECNADE and
support the current acceleration of the universe.

\section{\bf{Ho$\check{\text r}$ava-Lifshitz Gravity Theory and Interacting ECNADE Model}}

In the (3+1) dimensional Arnowitt-Deser-Misner formalism the full
metric is written as [51],

\begin{equation}
ds^{2}=-N^{2}dt^{2} + g_{ij}(dx^{i} + N^{i}dt)(dx^{j} + N^{j}dt)
\end{equation}

Under the detailed balance condition the full action condition of
Ho$\check{\text r}$ava-Lifshitz gravity [52] is given by,

\begin{eqnarray*}
S=\int dt d^{3}x \sqrt{g}N
\left[\frac{2}{\kappa^{2}}(K_{ij}K^{ij} - \lambda K^{2}) +
\frac{\kappa^{2}}{2 \omega^{4}}C_{ij}C^{ij}- \frac{\kappa^{2}\mu
\epsilon^{ijk}}{2 \omega^{2} \sqrt{g}}
R_{il}\nabla_{j}R^{l}_{k}\right.
\end{eqnarray*}
\begin{equation}
\left.- \frac{\kappa^{2}\mu^{2}}{8}R_{ij}R^{ij}+
\frac{\kappa^{2}\mu^{2}}{8(1-3\lambda)}\left(\frac{1 -4
\lambda}{4}R^{2} + \Lambda R - 3 \Lambda^{2} \right)\right]
\end{equation}

where

\begin{equation}
K_{ij}=\frac{1}{2N}(\dot{g}_{ij} - \nabla_{i}N_{j}-
\nabla_{j}N_{i})
\end{equation}

is the extrinsic curvature and

\begin{equation}
C^{ij}=\frac{\epsilon^{ikl}}{\sqrt{g}}\nabla_{k}(R^{j}_{i} -
\frac{1}{4}R \delta^{j}_{l})
\end{equation}

is known as Cotton tensor and the covariant derivatives are
defined with respect to the spatial metric $g_{ij}$.
$\epsilon^{ijk}$ is the totally antisymmetric unit tensor,
$\lambda$ is a dimensionless coupling constant and the variable
$\kappa$ , $\omega$ and $\mu$ are constants with mass dimensions
$-1,~ 0,~ 1$ respectively. Also $\Lambda$ is a positive constant,
which as usual is related to the cosmological constant in the IR
limit.\\

Now, in order to focus on cosmological frameworks, we impose the
so called projectability condition and use a FRW metric we get,

\begin{equation}
N=1,    g_{ij}=a^{2}(t)\gamma_{ij},      N^{i}=0
\end{equation}

with

\begin{equation}
\gamma_{ij} dx^{i} dx^{j}= \frac{dr^{2}}{1 - kr^{2}}+
r^{2}d\Omega^{2}_{2},
\end{equation}
where $k=0, -1, +1$ corresponding to flat, open and closed
respectively. By varying $N$  and $g_{ij}$, we obtain the
non-vanishing equations of motions:

\begin{equation}
H^{2}=\frac{\kappa^{2}}{6(3\lambda -1)} ~ \rho +
\frac{\kappa^{2}}{6(3\lambda -1)} \left[\frac{3 \kappa^{2} \mu^{2}
k^{2}}{8(3\lambda -1)a^{4}} + \frac{3 \kappa^{2} \mu^{2}
\Lambda^{2}}{8(3\lambda -1)}\right] - \frac{ \kappa^{4} \mu^{2}k
\Lambda }{8(3\lambda -1)^{2}a^{2}}
\end{equation}
and
\begin{equation}
\dot{H} + \frac{3}{2}H^{2}= - \frac{\kappa^{2}}{4(3\lambda -1)}~p
- \frac{\kappa^{2}}{4(3\lambda -1)} \left[\frac{ \kappa^{2}
\mu^{2} k^{2}}{8(3\lambda -1)a^{4}}- \frac{3 \kappa^{2} \mu^{2}
\Lambda^{2}}{8(3\lambda -1)}\right] - \frac{ \kappa^{4} \mu^{2}k
\Lambda }{16(3\lambda -1)^{2}a^{2}}
\end{equation}

The term proportional to $a^{-4}$ is the usual ``dark radiation",
present in Ho$\check{\text r}$ava-Lifshitz cosmology while the
constant term is just the explicit cosmological constant. For
$k=0$, there is no contribution from the higher order derivative
terms in the action. However for $k\ne 0$, their higher derivative
terms are significant for small volume i.e., for small $a$ and
become insignificant for large $a$, where it agrees with general
relativity. As a last step, requiring these expressions to
coincide the standard Friedmann equations, in units where $c=1$,

\begin{equation}
G_{c}=\frac{\kappa^{2}}{16 \pi(3\lambda -1)}
\end{equation}
and
\begin{equation}
\frac{ \kappa^{4} \mu^{2} \Lambda }{8(3\lambda -1)^{2}}=1
\end{equation}

where $G_{c}$ is the ``cosmological" Newton's constant. We mention
that in theories with Lorentz invariance breaking (such is
Ho$\check{\text r}$ava-Lifshitz one) the ``gravitational" Newton's
constant $G$, that is the one that is present in the gravitational
action, does not coincide with $G_{c}$, that is the one that is
present in Friedmann equations, where

\begin{equation}
G=\frac{\kappa^{2}}{32 \pi}
\end{equation}

as it can be straightforwardly read from the action. In the IR
$(\lambda=1)$ where Lorentz invariance is restored, $G_{c}=G$.
Using the above identifications, we can re-write the Friedmann
equations as,

\begin{equation}
H^{2} + \frac{k}{a^{2}}=\frac{8\pi G_{c}}{3}~(\rho_x+\rho_m) +
\frac{k^{2}}{2 \Lambda a^{4}} + \frac{\Lambda}{2}
\end{equation}
and
\begin{equation}
\dot{H} + \frac{3}{2}H^{2} + \frac{k}{2a^{2}}= -4 \pi
G_{c}(p_x+p_m) - \frac{k^{2}}{4 \Lambda a^{4}} +
\frac{3\Lambda}{4}
\end{equation}

Here, $\rho_{x}~~and~~p_x$ is the energy density and pressure of
the dark energy, $\rho_{m}~~and~~p_m$ is the energy density and
pressure of the matter respectively and $m_p^2=\frac{1}{8\pi
G_{c}}$ is modified Plank mass. Now equation (16) can be written
as,

\begin{equation}
1+\frac{k}{a^2 H^2} = \frac{\rho_{x}}{3m_{p}^2 H^2
}+\frac{\rho_{m}}{3m_{p}^2 H^2}+\frac{k^2}{2\Lambda a^4 H^2}
+\frac{\Lambda}{2 H^2}
\end{equation}

which gives,
\begin{equation}
1+\Omega_{k}=\Omega_{x}+\Omega_{m}+\frac{\Omega_{k}^{2}}{2\Omega_{\Lambda}}+\frac{\Omega_{\Lambda}}{2}
\end{equation}
where $\Omega_{k}, \Omega_{x}~and~\Omega_{m}$ are dimensionless
parameters defined as, $\Omega_{k}=\frac{k}{a^2 H^2},~~
\Omega_{x}=\frac{\rho_{x}}{3m_{p}^2 H^2},~~
\Omega_{\Lambda}=\frac{\Lambda}{H^2} ~and~
\Omega_{m}=\frac{\rho_{m}}{3m_{p}^2 H^2}$. \\

Now the energy density of the ECNADE is given in equation (4). If
$\beta=0 ~~and~~ \gamma=0$ then it is same as holographic dark
energy model. Now from (4) we can write,

\begin{equation}
\Omega_{x}=\frac{3n^2 m_{p}^2 T^2+\gamma \ln(m_{p}^2
T^{2})+\beta}{3m_{p}^2 H^2 T^4}
\end{equation}

Let us assume that there is an interaction between ENCADE and
pressureless cold dark matter and so the energy conservation
equations are,

\begin{equation}
\dot{\rho}_{x}+3H(\rho_{x}+p_{x})=-Q
\end{equation}
and
\begin{equation}
\dot{\rho}_{m}+3H\rho_{m}=Q
\end{equation}

where $Q=\Upsilon \rho_{x}$ and we choose
$\Upsilon=3b^2H(\frac{1+\Omega_{k}}{\Omega_{x}})$ with $b$ is a
coupling constant. So differentiating equation (4) w.r.t. `t' we
get,

\begin{equation}
\dot{\rho}_{x}=\frac{3\chi H}{a}\sqrt{\frac{2m_p^2
\Omega_{x}^{2}}{3n^2m_p^2+\gamma T^{-2}\ln(m_p^2 T^2)+\beta
T^{-2}}}
\end{equation}
where $\chi=-3n^2m_p^2 T^{-2}-2\gamma T^{-4}\ln(m_p^2 T^2)-2\beta
T^{-4}+\gamma T^{-4}$. Now $w_{ecnade}=\frac{p_{x}}{\rho_x}$ be
the equation of state of ECNADE. So using the conservation
equations we obtain the equation of state of ECNADE as,
\begin{equation}
w_{ecnade}=-1-b^2\left(\frac{1+\Omega_{k}}{\Omega_{x}}\right)-\frac{2\chi
T^{2}\Omega_{x}\sqrt{3m_{p}^{2}}}{3a\left(3n^2m_p^2+\gamma
T^{-2}\ln(m_p^2 T^2)+\beta T^{-2}\right)^{\frac{3}{2}}}
\end{equation}

\section{\bf{Correspondence of ECNADE Model with DBI-essence}}

There have been many works aimed at connecting the string theory
with inflation. While doing so, various ideas in string theory
based on the concept of branes have proved themselves fruitful.
One area which has been well explored in recent years, is
inflation driven by the open string sector through dynamical
Dp-branes. This is the so-called DBI (Dirac-Born-Infield)
inflation, which lies in a special class of K-inflation models.
Considering the dark energy scalar field is a Dirac-Born-Infeld
(DBI) scalar field, the action of the field can be written as
[4-11],
\begin{equation}
ds_{dbi}=-\int d^4x
a^3(t)\left[\digamma(\phi)\sqrt{1-\frac{\dot{\phi^{2}}}{\digamma(\phi)}}+V(\phi)-\digamma(\phi)\right]
\end{equation}
where $\digamma(\phi)$ is the tension and $V(\phi)$ is the
potential. From the above expression, the corresponding pressure
and the energy density of the scalar field becomes,

\begin{equation}
 p_{dbi} = \frac{\gamma-1}{\gamma}\digamma(\phi)-V(\phi)~~~and~~~ \rho_{dbi} = (\gamma-1)\digamma(\phi)+V(\phi)
\end{equation}
 where $\gamma$
 is reminiscent from the usual relativistic
Lorentz factor and is given by,

\begin{equation}
\gamma=\left(1-\frac{\dot{\phi^{2}}}{\digamma(\phi)}\right)^{-\frac{1}{2}}
\end{equation}

Thus the equation of state is given by,

\begin{equation}
w_{dbi}=\frac{\frac{\gamma-1}{\gamma}\digamma(\phi)-V(\phi)}{
(\gamma-1)\digamma(\phi)+V(\phi)}=\frac{(\gamma-1)\digamma(\phi)-\gamma
V(\phi)}{\gamma((\gamma-1)\digamma(\phi)+V(\phi))}
\end{equation}
\\
Now we compare the equation of states and energy densities of the
ECNADE and DBI-essence. Thus we get $w_{ecnade}=w_{dbi}$ and
$\rho_{x}=\rho_{dbi}$ which gives,
\begin{equation}
\frac{(\gamma-1)\digamma(\phi)-\gamma
V(\phi)}{\gamma((\gamma-1)\digamma(\phi)+V(\phi))}
=-1-b^2(\frac{1+\Omega_{k}}{\Omega_{x}})-\frac{2\chi}{3a}\left(\frac{\sqrt{\frac{3m_p^2
\Omega_{x}^2}{3n^2m_p^2+\gamma T^{-2}\ln(m_p^2 T^2)+\beta
T^{-2}}}}{3n^2m_p^2 T^{-2}+\gamma T^{-4}\ln(m_p^2 T^2)+\beta
T^{-4}}\right)
\end{equation}
and
\begin{equation}
\frac{3n^2 m_{p}^2}{T^{2}}+\frac{\gamma \ln(m_{p}^2
T^{2})+\beta}{T^4}=(\gamma-1)\digamma(\phi)+V(\phi)
\end{equation}

So from (27), (29) and (30) we get the analytical expressions of
$\digamma, \phi$ and $V$ as

$$
\digamma=-\frac{1}{3a(\gamma^2-1)T^{4}\Omega_{x}}\gamma\left[3ab^2(\beta+3m^2n^2T^2)(1+\Omega_{k})+2\sqrt{3}(\alpha-2\beta-3m^2n^2T^2)\Omega_{x}^{2}
\sqrt{\frac{m^2T^2}{\beta+3m^2n^2T^2+\alpha \ln[m^2T^2]}}\right.
$$
\begin{equation} \left.
+\alpha
\ln[m^2T^2]\left(3ab^2(1+\Omega_{k})-4\sqrt{3}\Omega_{x}^{2}
\sqrt{\frac{m^2T^2}{\beta+3m^2n^2T^2+\alpha
\ln[m^2T^2]}}~\right)\right]
\end{equation}

\begin{eqnarray*}
\phi=\int\left[\frac{1}{3a\gamma
T^{4}\Omega_{x}}\left\{-3ab^2(\beta+3m^2n^2T^2)(1+\Omega_{k})+2\sqrt{3}(-\alpha+2\beta+3m^2n^2T^2)\Omega_{x}^{2}
\sqrt{\frac{m^2T^2}{\beta+3m^2n^2T^2+\alpha
\ln[m^2T^2]}}\right.\right.
\end{eqnarray*}

\begin{equation}
\left.\left.~~+\alpha
\ln[m^2T^2]\left(-3ab^2(1+\Omega_{k})+4\sqrt{3}\Omega_{x}^{2}
\sqrt{\frac{m^2T^2}{\beta+3m^2n^2T^2+\alpha
\ln[m^2T^2]}}~\right)\right\}\right]^{\frac{1}{2}}dt
\end{equation}
and
$$
V=\frac{1}{3a(\gamma+1)T^{4}\Omega_{x}}\left[3a(\beta+3m^2n^2T^2)(b^2\gamma(1+\Omega_{k})+(1+\gamma)\Omega_{x})+2\sqrt{3}
(\alpha-2\beta-3m^2n^2T^2)\Omega_{x}^{2}\right.
$$
\begin{equation}
\left. \sqrt{\frac{m^2T^2}{\beta+3m^2n^2T^2+\alpha
\ln[m^2T^2]}}+\alpha
\ln[m^2T^2]\left(3a(b^2\gamma(1+\Omega_{k})+(1+\gamma)\Omega_{x})-4\sqrt{3}\Omega_{x}^{2}
\sqrt{\frac{m^2T^2}{\beta+3m^2n^2T^2+\alpha
\ln[m^2T^2]}}\right)\right]
\end{equation}

\section{\bf{Correspondence of ECNADE Model with Yang-Mills Dark Energy}}

Recent studies suggest that Yang-Mills field [12-16] can be
considered as a useful candidate to describe the dark energy as in
the normal scalar models the connection of field to particle
physics models has not been clear so far and the weak energy
condition cannot be violated by the field. In the effective Yang
Mills Condensate (YMC) dark energy model, the effective Yang-Mills
field Lagrangian is given by,
\begin{equation}
{\cal{L}}_{_{YMC}}=\frac{1}{2}bF(\ln\left|\frac{F}{k^2}\right|-1)
\end{equation}

where $k$ is the re-normalization scale of dimension of squared
mass, $F$ plays the role of the order parameter of the YMC where
$F$ is given by,
$F=-\frac{1}{2}F^{a}_{\mu\nu}F^{a\mu\nu}=E^2-B^2$. The pure
electric case we have, $B=0 ~~i.e. F = E^2$.\\

From the above Lagrangian we can derive the energy density and the
pressure of the YMC in the flat FRW spacetime as
\begin{equation}
\rho_{y}=\frac{1}{2}(y+1)b E^2
\end{equation}
and
\begin{equation}
p_{y}=\frac{1}{6}(y-3)b E^2
\end{equation}
where $y$ is defined as,
\begin{equation}
y=\ln\left|\frac{E^2}{k^2}\right|
\end{equation}
So the EoS of the YMC is given as,
\begin{equation}
w_{y}=\frac{p_y}{\rho_{y}}=\frac{y-3}{3y+3}
\end{equation}
Now we equate the EoS of the ECNADE and YMC i.e.
$w_{y}=w_{ECNADE}$ which gives,
\begin{equation}
\frac{y-3}{3y+3}=-1-b^2(\frac{1+\Omega_{k}}{\Omega_{x}})-\frac{2\chi}{3a}\left(\frac{\sqrt{\frac{3m_p^2
\Omega_{x}^2}{3n^2m_p^2+\gamma T^{-2}\ln(m_p^2 T^2)+\beta
T^{-2}}}}{3n^2m_p^2 T^{-2}+\gamma T^{-4}\ln(m_p^2 T^2)+\beta
T^{-4}}\right)
\end{equation}
which gives, {\small{\begin{equation}
y=\frac{-3ab^2(\beta+3m_p^2n^2T^2)(1+\Omega_{k})-2\sqrt{3}(\alpha-2\beta-3m_p^2n^2T^2)\Omega_{x}^{2}P+\alpha
\ln[m_p^2T^2]\left(-3ab^2(1+\Omega_{k})+4\sqrt{3}\Omega_{x}^2P\right)}{2\sqrt{3}P\Omega_{x}^{2}(\alpha-2\beta
-3m_p^2n^2T^2-2\alpha
\ln[m_p^2T^2])+a(3b^2(1+\Omega_{k})+4\Omega_{x})(\beta+3m_p^2n^2T^2+\alpha
\ln[m_p^2T^2])}
\end{equation}}}
where,
\begin{equation}
P=\sqrt{\frac{m_p^2T^2}{\beta+3m_p^2n^2T^2+\alpha \ln[m_p^2T^2]}}
\end{equation}

\section{\bf{Correspondence of ECNADE Model with Chameleon Field}}

Many of the dark energy models might give a large correction to
the Newton's law as the models are considered as a scalar field
rolling down a flat potential. But in our solar system this scalar
field becomes effectively massless. If this nearly massless scalar
field existed on Earth it should have been detected in local test
of equivalence principle and as a fifth force, unless the coupling
to normal matter was unnatural from a theoretical standpoint. So a
different approach is now being considered in general relativity,
where the quintessence scalar field is allowed to interact non
minimally with matter sector rather than the geometry and this
interaction is introduced through an interference term in the
action known as Chameleon scalar field [17-22].\\

The field equations of Ho$\check{\text r}$ava-Lifshitz gravity in
presence of chameleon field $\phi$ can be written as,
\begin{equation}
H^{2}+\frac{k}{a^2}=\frac{\rho_c f(\phi)}{3m_p^{2}}
+\frac{k^2}{2\Lambda a^4}+\frac{\Lambda}{2} +\frac{1}{2}
\dot{\phi}^2+V(\phi)=\frac{f(\phi)}{3m_p^{2}}(\rho_m+\rho_X)
+\frac{k^2}{2\Lambda a^4}+\frac{\Lambda}{2}
\end{equation}

\begin{equation}
\dot{H}+\frac{3}{2}H^{2}+\frac{k}{2a^2}=-\frac{p_c
f(\phi)}{2m_p^{2}} -\frac{k^2}{4\Lambda a^4}+\frac{3\Lambda}{2}
-\frac{1}{2}
\dot{\phi}^2+V(\phi)=-\frac{f(\phi)}{2m_p^{2}}(p_m+p_X)
-\frac{k^2}{4\Lambda a^4}+\frac{3\Lambda}{2}
\end{equation}
where $V(\phi)$ is the relevant potential and $f(\phi)$ is any
analytical function of $\phi$. Also
$\rho_X=\frac{3m_p^2}{f(\phi)}[\frac{1}{2} \dot{\phi}^2+V(\phi)]$
and $p_X=\frac{2m_p^2}{f(\phi)}[\frac{1}{2}
\dot{\phi}^2-V(\phi)]$.\\

So the EoS of the Chameleon field is given as,
\begin{equation}
w_{X}=\frac{p_X}{\rho_{X}}=\frac{2(\frac{1}{2}
\dot{\phi}^2+V(\phi))}{3(\frac{1}{2} \dot{\phi}^2-V(\phi))}
\end{equation}
Now we equate the EoS and energy density of the Chameleon Field
and ECNADE i.e. $w_{X}=w_{ECNADE}$ and $\rho_X=\rho_{ECNADE}$
which gives,
\begin{equation}
\frac{2(\frac{1}{2} \dot{\phi}^2+V(\phi))}{3(\frac{1}{2}
\dot{\phi}^2-V(\phi))}=-1-b^2(\frac{1+\Omega_{k}}{\Omega_{x}})-\frac{2\chi}{3a}\left(\frac{\sqrt{\frac{3m_p^2
\Omega_{x}^2}{3n^2m_p^2+\gamma T^{-2}\ln(m_p^2 T^2)+\beta
T^{-2}}}}{3n^2m_p^2 T^{-2}+\gamma T^{-4}\ln(m_p^2 T^2)+\beta
T^{-4}}\right)
\end{equation}
and
\begin{equation}
\frac{3m_p^2}{f(\phi)}[\frac{1}{2}
\dot{\phi}^2+V(\phi)]=\frac{3n^2 m_{p}^2}{T^{2}}+\frac{\gamma
\ln(m_{p}^2 T^{2})+\beta}{T^4}
\end{equation}

Solving the above two equations we get the expressions of $f$ and
$V$ as, {\small{\begin{equation} f=-\frac{6a
m^2T^4\dot{\phi}^2\Omega_{X}}{(a(\beta+3m_p^2n^2T^2)(3b^2(1+\Omega_{k})+\Omega_{X})+2\sqrt{3}(\alpha-2\beta-3m_p^2n^2T^2)
\Omega_{x}^{2}P+\alpha\ln[m_p^2T^2]\left(a(3b^2(1+\Omega_{k})+\Omega_{X})-4\sqrt{3}\Omega_{x}^2P\right))}
\end{equation}}}
and {\small{\begin{equation}
V=-\frac{\dot{\phi}^{2}((a(\beta+3m_p^2n^2T^2)(3b^2(1+\Omega_{k})+5\Omega_{X})+2\sqrt{3}(\alpha-2\beta-3m_p^2n^2T^2)
\Omega_{x}^{2}P+\alpha\ln[m_p^2T^2]\left(a(3b^2(1+\Omega_{k})+5\Omega_{X})-4\sqrt{3}\Omega_{x}^2P\right)))}
{2(a(\beta+3m_p^2n^2T^2)(3b^2(1+\Omega_{k})+\Omega_{X})+2\sqrt{3}(\alpha-2\beta-3m_p^2n^2T^2)
\Omega_{x}^{2}P+\alpha\ln[m_p^2T^2]\left(a(3b^2(1+\Omega_{k})+\Omega_{X})-4\sqrt{3}\Omega_{x}^2P\right))}
\end{equation}}}
where,
\begin{equation}
P=\sqrt{\frac{m_p^2T^2}{\beta+3m_p^2n^2T^2+\alpha \ln[m_p^2T^2]}}
\end{equation}

\section{\bf{Correspondence of ECNADE Model with Non-linear Electrodynamics Field}}

Recently a new approach has been taken to avoid the cosmic
singularity through a nonlinear extension of the Maxwell
electromagnetic theory. Another interesting feature can be viewed
that an exact regular black hole solution has been recently
obtained proposing Einstein-dual nonlinear electrodynamics. Exact
solutions of the Einstein's field equations coupled with nonlinear
electrodynamics (NLED) reveal an acceptable nonlinear effect in
strong gravitational and magnetic fields. Also the General
Relativity (GR) coupled with NLED effects can explain the
primordial inflation.\\

The Lagrangian density for free fields in the Maxwell
electrodynamics may be written as [23-25],
\begin{equation}
{\cal{L_M}}=-\frac{1}{4\mu}F^{\mu\nu}F_{\mu\nu}
\end{equation}
where $F^{\mu\nu}$ is the electromagnetic field strength tensor
and $\mu$ is the magnetic permeability.\\

Here we consider the generalization of Maxwell electro-magnetic
Lagrangian up to the second order terms of the fields as

\begin{equation}
{\cal L}=-\frac{1}{4\mu_{0}}~F+\omega F^{2}+\eta F^{*2}
\end{equation}

where $\omega$ and $\eta$ are arbitrary constants,

\begin{equation}
F^{*}\equiv F^{*}_{\mu\nu}F^{\mu\nu}
\end{equation}

and $F^{*}_{\mu\nu}$ is the dual of $F_{\mu\nu}$. Now we consider
the homogeneous electric field $E$ in plasma gives rise to an
electric current of charged particles and then rapidly decays. So
the squared magnetic field $B^{2}$ dominates over $E^{2}$, i.e.,
in this case, $E^{2}\approx 0$ and hence $F=2B^{2}$. So $F$ is now
only the function of magnetic field.\\

Now the pressure and energy density of the Non-linear
Electrodynamics Field is,
\begin{equation}
p_{_{NE}}=\frac{B^2}{6\mu}(1-40\mu \omega B^{2})
\end{equation}
and
\begin{equation}
\rho_{_{NE}}=\frac{B^2}{2\mu}(1-8\mu \omega B^{2})
\end{equation}
The weak condition($\rho>0$) is obeyed for,
$B<\frac{1}{2\sqrt{2\mu\omega}}$ and pressure will be negative
when $B>\frac{1}{2\sqrt{10\mu\omega}}$. The magnetic field
generates dark energy if strong energy condition is violated i.e.,
$\rho_{B}+3p_{B}<0$, i.e., if $B>\frac{1}{2\sqrt{6\mu\omega}}$.\\

So the EoS of the Nonlinear Electrodynamics Field is given as,
\begin{equation}
w_{NE}=\frac{p_{NE}}{\rho_{NE}}=\frac{\frac{B^2}{6\mu}(1-40\mu
\omega B^{2})}{\frac{B^2}{2\mu}(1-8\mu \omega
B^{2})}=\frac{1-40\mu \omega B^{2}}{3(1-8\mu \omega B^{2})}
\end{equation}
Now we equate the EoS  of the Nonlinear Electrodynamics Field and
ECNADE i.e. $w_{NE}=w_{ECNADE}$ gives,
\begin{equation}
\frac{(1-40\mu \omega B^{2})}{3(1-8\mu \omega
B^{2})}=-1-b^2(\frac{1+\Omega_{k}}{\Omega_{x}})-\frac{2\chi}{3a}\left(\frac{\sqrt{\frac{3m_p^2
\Omega_{x}^2}{3n^2m_p^2+\gamma T^{-2}ln(m_p^2 T^2)+\beta
T^{-2}}}}{3n^2m_p^2 T^{-2}+\gamma T^{-4}ln(m_p^2 T^2)+\beta
T^{-4}}\right)
\end{equation}
From where we get the expression of magnetic field as,
{\small{\begin{equation}
B=\sqrt{\frac{a(\beta+3m_p^2n^2T^2)(3b^2(1+\Omega_{k})+4\Omega_{X})+2\sqrt{3}(\alpha-2\beta-3m_p^2n^2T^2)
\Omega_{x}^{2}P+\alpha\ln[m_p^2T^2]\left(a(3b^2(1+\Omega_{k})+4\Omega_{X})-4\sqrt{3}\Omega_{x}^2P\right)
}{8\mu\omega(a(\beta+3m_p^2n^2T^2)(3b^2(1+\Omega_{k})+8\Omega_{X})+2\sqrt{3}(\alpha-2\beta-3m_p^2n^2T^2)
\Omega_{x}^{2}P+\alpha\ln[m_p^2T^2]\left(a(3b^2(1+\Omega_{k})+8\Omega_{X})-4\sqrt{3}\Omega_{x}^2P\right))}}
\end{equation}}}

where,
\begin{equation}
P=\sqrt{\frac{m_p^2T^2}{\beta+3m_p^2n^2T^2+\alpha \ln[m_p^2T^2]}}
\end{equation}

\section{\bf{Correspondence of ECNADE Model with Hessence Dark Energy}}

Wei et al [26] proposed a novel non-canonical complex scalar field
named ``hessence" which play the role of quintom. In the hessence
model the so called internal motion $\dot{\theta}$ where $\theta$
is the internal degree of freedom of hessence plays a Phantom like
role and the Phantom divide transitions is also possible.\\

Now the pressure and energy density of the hessence dark energy is
given by [27, 28]

\begin{equation}
p_{_{X}}=\frac{1}{2}(\dot{\phi}^{2}-\phi^{2}\dot{\theta}^{2})-V(\phi)
\end{equation}
and
\begin{equation}
\rho_{_{X}}=\frac{1}{2}(\dot{\phi}^{2}-\phi^{2}\dot{\theta}^{2})+V(\phi)
\end{equation}

where $Q=a ^{3}\phi^{2}\dot{\theta}= \text{constant}$ is the total
conserved charge. So the EoS of the  is given as,
\begin{equation}
w_{X}=\frac{p_X}{\rho_{X}}=\frac{\frac{1}{2}(\dot\phi^{2}-\frac{Q^{2}}
{a^{6}\phi^{2}})-V(\phi)}{\frac{1}{2}(\dot\phi^{2}-\frac{Q^{2}}{a^{6}\phi^{2}})+V(\phi)}
\end{equation}
Now we equate the EoS and energy density of the and ECNADE i.e.
$w_{X}=w_{ECNADE}$ and $\rho_X=\rho_{ECNADE}$ which gives,
\begin{equation}
\frac{p_X}{\rho_{X}}=\frac{\frac{1}{2}(\dot\phi^{2}-\frac{Q^{2}}
{a^{6}\phi^{2}})-V(\phi)}{\frac{1}{2}(\dot\phi^{2}-\frac{Q^{2}}{a^{6}\phi^{2}})+V(\phi)}
=-1-b^2(\frac{1+\Omega_{k}}{\Omega_{x}})-\frac{2\chi}{3a}\left(\frac{\sqrt{\frac{3m_p^2
\Omega_{x}^2}{3n^2m_p^2+\gamma T^{-2}ln(m_p^2 T^2)+\beta
T^{-2}}}}{3n^2m_p^2 T^{-2}+\gamma T^{-4}ln(m_p^2 T^2)+\beta
T^{-4}}\right)
\end{equation}
and
\begin{equation}
\frac{1}{2}(\dot{\phi}^{2}-\frac{Q^{2}}{a^{6}\phi^{2}})+V(\phi)=\frac{3n^2
m_{p}^2}{T^{2}}+\frac{\gamma ln(m_{p}^2 T^{2})+\beta}{T^4}
\end{equation}

which gives,
 $ V(\phi)=\frac{1}{6 a T^{4}\Omega_{X}}\left[3a(\beta+3m_p^2n^2T^2)(b^2(1+\Omega_{k})+2\Omega_{X})
+2\sqrt{3}(\alpha-2\beta-3m_p^2n^2T^2) \Omega_{X}^{2}P \right. $

\begin{equation}
\left.
+\alpha\ln[m_p^2T^2]\left(3a(b^2(1+\Omega_{k})+2\Omega_{X})-4\sqrt{3}\Omega_{X}^2P\right)\right]
\end{equation}
and
\begin{eqnarray*}
\phi=\int\frac{1}{\sqrt{3}}\left[\sqrt{\frac{1}{a^6
T^{4}\phi^{2}\Omega_{X}}}\left[\sqrt{-3a^{6}b^2(\beta+3m_p^2n^2T^2)
\phi^2(1+\Omega_{k})+3Q^2 T^4\Omega_{X}}\right. \right.
\end{eqnarray*}

\begin{equation}
 \left.\left.+ \overline{2\sqrt{3}a^5(-\alpha+2\beta+3m_p^2n^2T^2)\phi^2
\Omega_{X}^{2}P+a^5\alpha\phi^2\ln[m_p^2T^2]\left(-3ab^2(1+\Omega_{k})+4\sqrt{3}\Omega_{X}^2P\right)}\right]\right]dt
\end{equation}
where,
\begin{equation}
P=\sqrt{\frac{m_p^2T^2}{\beta+3m_p^2n^2T^2+\alpha \ln[m_p^2T^2]}}
\end{equation}

\section{\normalsize\bf{Discussions}}

Several authors have extensively discussed about ECHDE [52-55] and
ECNADE [56-59] model to explain recent acceleration in different
context using different dark energy models. In this work, we have
considered the entropy-corrected new agegraphic dark energy
(ECNADE) model is applied in a universe governed by a modified
gravity model named  Ho$\check{\text r}$ava-Lifshitz gravity in
FRW universe to understand the behaviour of the potentials when a
correspondence  between the ECNADE and other dark energies have
been considered. Since the new agegraphic dark energy (NADE) model
in the framework of quantum gravity is being used as a source to
probe the nature of dark energy, so we have used the logarithmic
corrected version of NADE which is usually motivated from the
important feature of LQG to understand tghe nature of the universe
in very early epoch. The interaction between ECNADE and dark
matter have been considered. Choosing some particular form of
interaction term, we have derived the equation of state for
ECNADE. Recent studies reveals that the scalar field models are
very effective as a candidate of dark energy, so we use these
models according to the evolutionary behavior of the interacting
ECNADE. So we compare the equation of state of different scalar
field models with the ECNADE. We have discussed the correspondence
between ECNADE and other dark energy models such as DBI-essence,
Yang-Mills dark energy, Chameleon field, Non-linear
electrodynamics field and hessence dark energy in the context of
Ho$\check{\text r}$ava-Lifshitz gravity and reconstructed the
potentials and the dynamics of the scalar field theory which
describe the ECNADE.\\\\

{\bf Acknowledgement:}\\

The authors are thankful to IUCAA, Pune, India for warm
hospitality where part of the work was carried out.\\\\

{\bf References:}\\\

 [1] A. G. Riess et al,  \textit{Astron. J.} \textbf{116} 1009 (1998).\\\

 [2] D. N. Spergel \textit{Astrophys. J. Suppl.} \textbf{148} 175 (2003).\\\

 [3] S. Perlmutter et al, \textit{Astrophys. J.} \textbf{517} 565 (1999).\\\

 [4] J. E. Lidsey and I. Huston,   \textit{JCAP} \textbf{0707} 002 (2007).\\\

 [5] W. H. Kinney and K. Tzirakis, \textit{Phys. Rev. D} \textbf{77} 103517 (2008).\\\

 [6] J. Martin and M. Yamaguchi,  \textit{Phys. Rev. D} \textbf{77} 123508 (2008). \\\

 [7] M. Spali$\acute{n}$ski, \textit{JCAP} \textbf{017} 05 (2007); \textit{Phys. Lett. B} \textbf{650} 313 (2007).\\\

 [8] X. Chen, M. X. Huang, S. Kachru and G. Shiu,  \textit{JCAP} \textbf{0701} 002 (2007).\\\

 [9] S. Kecskemeti, J. Maiden, G. Shiu and B. Underwood, \textit{JHEP} \textbf{0609} 076 (2006).\\\

 [10] M. X. Huang, G. Shiu and B. Underwood,  \textit{Phys. Rev. D} \textbf{77} 023511 (2008).\\\

 [11] D. Seery and J. E. Lidsey,  \textit{Phys. Rev. D} \textbf{75} 043505 (2007).\\\

 [12] Y. Zhang, T. Y. Xia and W. Zhao,  \textit{Class. Quant. Grav.} \textbf{24} 3309 (2007).\\\

 [13] T. Y. Xia and Y. Zhang, \textit{Phys. Lett. B} \textbf{656} 19 (2007).\\\

 [14] M. Tong, Y. Zhang and T. Xia, \textit{Int. J. Mod. Phys. D} \textbf{18} 797 (2009).\\\

 [15] W. Zhao, \textit{Int. J. Mod. Phys. D}\textbf{17} 1245 (2008); W. Zhao and Y. Zhang, \textit{Class. Quant. Grav.} \textbf{23}
3405 (2006); W. Zhao, \textit{Astron. Astrophys. }\textbf{9} 874
(2009); W. Zhao and D. Xu, \textit{Int. J. Mod. Phys. D}
\textbf{16} 1735 (2007);
Y. Zhang, T. Y. Xia , and W. Zhao, \textit{Class. Quant. Grav.}\textbf{24} 3309 (2007).\\\

 [16] Z. Yang, \textit{Chin. Phys. Lett.} \textbf{21} 1183
(2004).\\\

 [17] A. -C. Davis, C. A. O. Schelpe and D. J. Shaw, \textit{Phys. Rev. D} \textbf{80} 064016 (2009).\\\

 [18] S. Chakraborty and U. Debnath, \textit{IJMPA} \textbf{25} 4691 (2010).\\\

 [19] N. Banerjee et al, \textit{Pramana }\textbf{74} L481 (2010); N. Banerjee et al, \textit{Phys. Rev. D} \textbf{78} 043512 (2008).\\\

 [20] Ph. Brax, \textit{arXiv:}\textbf{0410103} [astro-ph] (2004);  Proceedings of the ``Phi in the Sky"
 conference ``The Quest for Cosmological Scalar Fields", doi:10.1063/1.1835177, Vol. 736, p. 105-110,
 8-10 July 2004, Porto, Portugual.\\\

 [21] H. Wei and R. -G. Cai, \textit{Phys. Rev. D}\textbf{71} 043504 (2005).\\\

 [22] D. F. Mota and D. J. Shaw, \textit{Phys. Rev. Lett.} \textbf{97}  151102 (2006).\\\

 [23] R. Garc´ýa-Salcedo et al, \textit{arXiv:}\textbf{1006.2276} [gr-qc] (2010).\\\

 [24] V. A. De Lorenci et al,  \textit{Phys. Rev. D} \textbf{65} 063501 (2002).\\\

 [25] C. S. Camara et al, \textit{Phys. Rev. D} \textbf{69} 103504 (2004).\\\

 [26] H. Wei, R. -G. Cai and D. F. Zhang,  \textit{Class. Quant. Grav.} \textbf{22} 3189 (2005);
 H. Wei and S. N. Zhang, \textit{Phys. Rev. D}\textbf{76}  063005 (2007);
 H. Wei, N. Tang and S. N. Zhang, \textit{Phys. Rev. D }\textbf{75}  043009 (2007).\\\

 [27] M. Alimohammadi and H. M. Sadjadi, \textit{Phys. Rev. D}\textbf{73} 083527 (2006).\\\

 [28] S. Chakraborty and U. Debnath, \textit{Int. J. Mod. Phys. D} \textbf{19} 2071 (2010).\\\

 [29] S. Nojiri and S. D. Odintsov  \textit{Int. J. Geom. Meth. Mod. Phys.} \textbf{4}  115 (2007).\\\

 [30] S. Nojiri and S. D. Odintsov,  \textit{arXiv:} \textbf{0807.0685}  [hep-th] (2008);
 {\it Problems of Modern Theoretical Physics}, A Volume in honour of Prof. I. L. Buchbinder in the occasion
 of his 60th birthday, p.266-285, TSPU Publishing, Tomsk.\\\

 [31] P. Ho$\check{\text r}$ava,  \textit{JHEP} \textbf{0903} 020 (2009).\\\

 [32] P. Ho$\check{\text r}$ava, \textit{Phys. Rev. D} \textbf{79} 084008 (2009).\\\

 [33] P. Ho$\check{\text r}$ava, {\it Phys. Rev. Lett.} {\bf 102} 161301 (2009).\\\

 [34] P. Ho$\check{\text r}$ava, {\it Phys. Lett. B} {\bf 694} 172 (2010).\\\

 [35] E. M. Lifshitz,  \textit{Zh. Eksp. Teor. Fiz.} \textbf{11} 255 (1949). \\\

 [36] R. G. Cai et al, \textit{Phys. Rev. D} \textbf{80} 041501 (2009); M. R. Setare and M. Jamil,  \textit{JCAP} \textbf{02} 010 (2010).\\\

 [37] G. Calcagni, \textit{JHEP} \textbf{0909} 112 (2009).\\\

 [38] H. Lu et al, \textit{Phys. Rev. Lett.} \textbf{103} 091301 (2009).\\\

 [39] G. t Hooft, \textbf{9310026}  [gr-qc] (2009); L. Susskind, \textit{J. Math. Phys.} \textbf {36} 6377
     (1995).\\\

 [40] S. W. Hawking, \textit{Commun. Math. Phys.} \textbf{43} 199 (1975); S. W. Hawking, \textit{Phys. Rev. D} \textbf{13} 191 (1976);
 J. D. Bekenstein, \textit{Phys. Rev. D} \textbf{23} 287 (1981).\\\

  [41] A. Cohen, D. Kaplan and A. Nelson, \textit{Phys. Rev. Lett.} \textbf{82} 4971
 (1999).\\\

  [42]  P. Ho$\check{\text r}$ava and D. Minic, \textit{Phys. Rev. Lett.} \textbf{85} 1610 (2000);
  S. D. Thomas, \textit{Phys. Rev. Lett.} \textbf{89} 081301 (2002).\\\

  [43] S. D. H. Hsu, \textit{Phys. Lett. B} \textbf{594}, 13 (2004).\\\

  [44] K. A. Meissner, \textit{Class. Quantum Grav.} \textbf{21} 5245 (2004).\\\

  [45] H. Wei, \textit{Commun. Theor. Phys.} \textbf{52} 743 (2009).\\\

  [46] R. -G. Cai,  \textit{Phys. Lett. B} \textbf{657} 228 (2007).\\\

 [47] H. Wei and R. -G. Cai,  \textit{Phys. Lett. B} \textbf{660} 113 (2008).\\\

 [48] F. Karolyhazy, \textit{Nuovo. Cim. A} \textbf{42} 390
(1966).\\\

 [49] M. Maziashvili, \textit{Int. J. Mod. Phys. D} \textbf{16} 1531 (2007); M. Maziashvili, \textit{Phys. Lett. B} \textbf{652} 165
  (2007).\\\

 [50] H. Wei, \textit{Commun. Theor. Phys.} \textbf{52} 743 (2009);
 M. Jamil and M. U. Farooq, \textit{JCAP} \textbf{03} 001 (2010).\\\

[51] R. L. Arnowitt, S. Deser and C. W. Misner,
\textit{Gravitation: an introduction to current research",
 Louis Witten ed. (Wiley 1962), chapter 7, pp 227--265} (2004).\\\

 [52] W. Hao, \textit{Commun. Theor. Phys.} \textbf{52} 743 (2009).\\\

 [53] M. R. Setare and M. Jamil, \textit{Europhys. Lett.} \textbf{92} 49003 (2010).\\\

 [54] A. K. Mohammadi, and M. Malekjani,  \textit{arXiv:} \textbf{1004.1720v2} [gr-qc] (2011).\\\

 [55] E. Ebrahimi and A. Sheykhi,  \textit{arXiv:} \textbf{1011.5005v2} [hep-th] (2011).\\\

 [56] M. Jamil and A. Sheykhi, \textit{Int. J. Theor. Phys.} \textbf{50} 625 (2011).\\\

 [57] M. Malekjani and A. K. Mohammadi, \textit{arXiv:} \textbf{1004.1017v2} [gr-qc] (2010).\\\

 [58] K. Karami et al, \textit{Europhys. Lett.} \textbf{93} 69001 (2011); K. Karami et al,
 \textit{Gen. Rel. Grav.} \textbf{43} 27 (2011); K. Karami and A. Sorouri, \textit{Phys. Scr.} \textbf{82} 025901 (2010).\\\

 [59] M. U. Farooq, M. Jamil and M. A. Rashid, \textit{Int. J. Theor. Phys.} \textbf{49} 2278 (2010).\\\

\end{document}